\documentclass[conference]{IEEEtran}
\IEEEoverridecommandlockouts
\usepackage{cite}
\usepackage{amsmath,amssymb,amsfonts}
\usepackage{multirow,tabularx}   
\usepackage{algorithmic}
\usepackage{graphicx}
\usepackage{steinmetz}
\usepackage{textcomp}
\usepackage{xcolor}
\def\BibTeX{{\rm B\kern-.05em{\sc i\kern-.025em b}\kern-.08em
    T\kern-.1667em\lower.7ex\hbox{E}\kern-.125emX}}
    
\usepackage{comment}     
\begin{document}

\title{A Novel Grid-forming Voltage Control Strategy for Supplying Unbalanced Microgrid Loads Using Inverter-based Resources
%{\footnotesize \textsuperscript{*}Note: }
\thanks{This  research  is  supported  by  the  U.S.  Department  of  Energy’s  Office  of Energy  Efficiency  and  Renewable  Energy  (EERE)  under  the  Solar  EnergyTechnologies Office Award Number DE-EE0008770.}
}

\author{\IEEEauthorblockN{Bei Xu, Victor Paduani, Hui Yu, David Lubkeman, Ning Lu}
\IEEEauthorblockA{
\textit{FREEDM Systems Center}, \textit{Department of Electrical and Computer Engineering} \\
North Carolina State University, Raleigh, NC \\
\{bxu8, vdaldeg, hyu11, dllubkem, nlu2\}@ncsu.edu}
%\and
}

\maketitle

\begin{abstract}
This paper presents a novel grid-forming voltage control strategy for a battery energy storage system to maintain balanced three-phase output voltages when serving unbalanced loads.
A stationary reference frame ($\alpha \beta$) based control scheme is proposed to regulate positive-sequence and negative-sequence voltages. Compared with the conventional rotating reference frame ($dq$) based control scheme, the proposed scheme shows better dynamic performance. Then, we analyze the system zero-sequence network and propose adding a grounding transformer to the Y/Yg connected output transformer to further reduce the voltage unbalance at the point of common coupling. The simulation results demonstrate the effective performance of the proposed voltage control scheme. Based on the results, the power-voltage unbalance curve is derived for different output transformer configurations to establish the relationship between the power unbalance limit and the voltage unbalance limit for microgrid power scheduling.

\end{abstract}


\section{Introduction}
Grid-forming inverter-based resources (IBRs) are critical for regulating voltage and frequency while responding to frequent load fluctuations in off-grid, microgrid operation. 
One distinct operational requirement in distribution system operation is to serve unbalanced loads.  
%When serving residential/commercial microgrids
When delivering unbalanced three-phase load currents, the non-zero internal impedance of an IBR leads to unbalanced voltages at the point of common coupling (PCC). However, severely unbalanced system voltage can cause damages or deteriorated operation to many voltage-sensitive loads, for example, three-phase motor loads \cite{MGprotection_2017}. 

In the literature, there are two general considerations when designing grid-forming IBRs for balancing system voltage: topology and control. Inverter topology design has three developed typologies: three-phase three-leg inverter with split dc-link capacitors, three-phase four-leg inverter\cite{four_leg_vechiu2009}, and three-phase combined inverter\cite{three-phase_combined_inverters2009}. The second topology and corresponding control strategies are complex and not suitable for the system that needs an isolation transformer. The third topology consists of three single-phase inverters, which is mostly used in residential low-power applications. Thus, in this paper, we focus on the three-phase, three-leg inverter topology for grid-forming voltage regulation because this topology is commonly used by manufacturers and is suitable for high-power applications.

There are two commonly used reference frames for inverter controller design: $dq$ rotating (synchronous) reference frame (RRF)\cite{book_tleis2007power,dq_2015ECCE_PR} and $\alpha\beta$ stationary reference frame (SRF) \cite{albe_2012autonomous,albe_DG2015APEC}. 
In general, RRF-based control scheme uses two pairs of proportional integral (PI) controllers, which situate in two reference frames while rotating in the opposite directions, to regulate positive-sequence (PS) and negative-sequence (NS) components, respectively\cite{book_tleis2007power}. 
%In \cite{dq_cao2017full}, Cao \emph{et al}. applied adaptive adjustment of the compensation for improving virtual power decoupling. 
However, the RRF-based control strategies need band-stop filters to acquire the DC components of the measurement signals. Adding filers introduce delays that can worsen the dynamic performance of voltage regulation. 
In \cite{dq_2015ECCE_PR},  Cai \emph{et al}. applied the proportional resonant (PR) controller for NS voltage regulation to avoid the delay caused by the added filter, but it requires high accuracy for the decomposition of PS and NS components.

SRF-based control scheme requires one pair of PR controllers for voltage regulation in AC domain. In \cite{albe_2012autonomous,albe_DG2015APEC}, the voltage reference is generated by the droop control and an unbalance compensator. The power used in droop control is calculated using PS components so that the unbalance compensator is applied for the reference generation to regulate the NS voltage. However, the proposed control scheme in this paper can achieve NS regulation without the need for the unbalance compensator.
Moreover, in the previous works, no output transformer is connected with the inverter. Thus, zero-sequence (ZS) network components are not analyzed and ZS voltage regulation is not considered.

In this paper, we propose an $\alpha\beta$ SRF-based control structure for a battery-energy-storage system (BESS) to regulate the PS and NS voltage while using a grounding transformer (GT) to reduce the ZS voltage. Our contributions are threefold. \emph{First}, we clearly demonstrated the advantages of the SRF-based inner voltage control scheme for PS and NS voltage regulation over the RRF-based control scheme. 
%A SRF-based inverter controller is developed without the need for the decomposition of the positive- and negative-sequence components.
\emph{Second}, we demonstrated that ZS voltage caused by unbalanced loads can be effectively reduced when adding a GT at the PCC by analyzing the ZS network components. 
\emph{Third}, we derived the relationship between power and voltage unbalance and proposed a performance metric for regulating power unbalance to meet inverter voltage unbalance requirements.  
%The proposed control scheme is validated by an IEEE 123-bus testbed consisting of a 2-MVA BESS through real-time simulation. 

\section{Voltage Control Strategy} \label{control}
\begin{figure}[!t]
 \vspace{-8pt}
  \centering
	\includegraphics[width=0.48\textwidth]{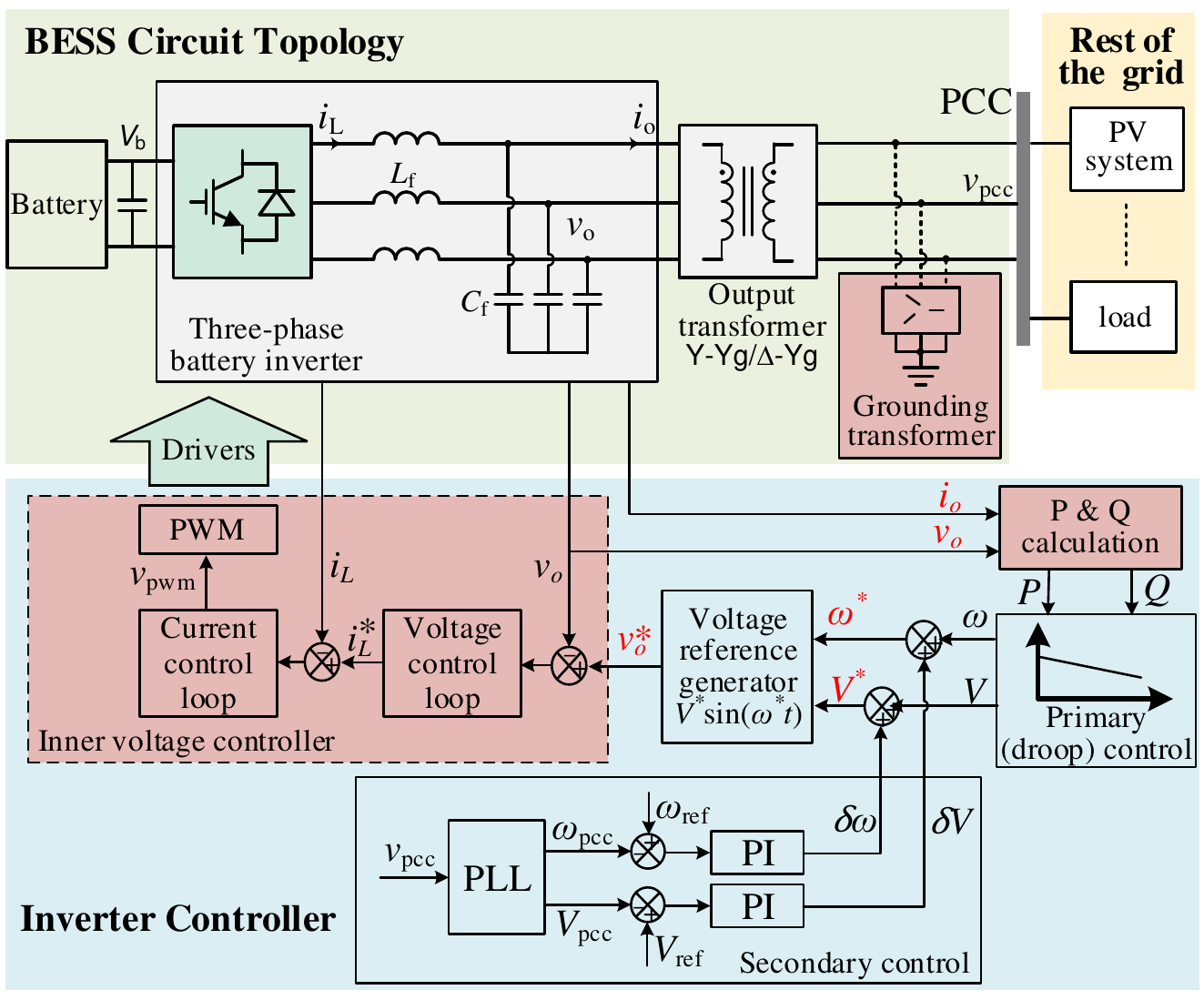}
	\vspace{-4pt}
	\caption{Topology and control structure of a three-phase grid-forming BESS.}
	\label{fig:ciruit}
	\vspace{-5pt}
\end{figure}

As shown in Fig. \ref{fig:ciruit}, a grid-forming BESS consists of a three-phase inverter, an LC filter, and a Y-Yg (or $\Delta$-Yg) isolation transformer for connecting the BESS to the main grid. 
In\cite{standard_2010hierarchical}, Guerrero \emph{et al}. proposed a commonly used droop-based hierarchical control structure, consisting of inner current and voltage control loops, primary (droop) control, and secondary control. This structure allows the seamless transition between the off-grid mode and the grid-connected mode without changing the control algorithm. When working in the off-grid mode, the grid-forming BESS inverter operates as a voltage source with two main functions: regulating PCC voltage, $v_\mathrm{pcc}$, and setting system frequency, $f$. 

Our contributions to this topology and control scheme are highlighted in red in Fig. \ref{fig:ciruit}. 
First, we designed an $\alpha\beta$ SRF-based inverter controller. The voltage reference of the inner voltage controller, $v_\mathrm{o}^*$, is generated based on the conventional droop and secondary control without the need for the decomposition of the PS and NS components. The inner voltage controller design for PS and NS voltage regulation is described in Section \ref{design} in detail. Second, we added a GT in the circuit to mitigate the impact of ZS currents on voltage regulation for keeping $v_\mathrm{pcc}$ balanced when serving highly unbalanced single-phase loads.

\subsection{Positive- and Negative- Sequence Voltage Representation}\label{mechanism}

Unbalanced three-phase phasors can be decomposed into PS, NS, and ZS phasors\cite{book_tleis2007power}. 
%According to the superimpose theorem, the unsymmetrical output voltage in unbalanced operations is the sum of response when positive-, negative-, and zero sequence currents are applied to the inverter, respectively. The zero sequence component is related to the transformer configuration, which is analyzed in detail in subsection \ref{impact}.
The PS and NS voltages are expressed as:

\begin{equation}
    \begin{bmatrix}
        v_{a}^+ \\
        v_{b}^+ \\
        v_{c}^+
    \end{bmatrix}
    =
    \begin{bmatrix}
        V_{m}^+ \cos(\omega t + \varphi^+) \\
        V_{m}^+ \cos(\omega t - 2 \pi / 3 + \varphi^+) \\
        V_{m}^+ \cos(\omega t + 2 \pi / 3 + \varphi^+)
    \end{bmatrix}
\end{equation}

\begin{equation}
    \begin{bmatrix}
        v_{a}^- \\
        v_{b}^- \\
        v_{c}^-
    \end{bmatrix}
    =
    \begin{bmatrix}
        V_{m}^- \cos(\omega t + \varphi^-) \\
        V_{m}^- \cos(\omega t + 2 \pi / 3 + \varphi^-) \\
        V_{m}^- \cos(\omega t - 2 \pi / 3 + \varphi^-)
    \end{bmatrix}
\end{equation}
where $V_{m}^+$ and $V_{m}^-$ are the amplitude of PS and NS voltage, respectively; $\varphi^+$ and $\varphi^-$ are the phase of PS and NS voltage, respectively. 
By using the Clarke transformation matrix $T_1$, the voltages in SRF ($\alpha\beta$-coordinate) are obtained:

\begin{subequations}\label{eq:albe}
\centering
\vspace{+3pt}
    \begin{minipage}[c]{0.48\textwidth}
    \begin{equation}
    \begin{bmatrix}
        v_{\alpha}^+ \\
        v_{\beta}^+ 
    \end{bmatrix}
    = T_1
    \begin{bmatrix}
        v_a^+ \\
        v_b^+ \\
        v_c^+ 
    \end{bmatrix}
    = 
    \begin{bmatrix}
    V_{m}^+ \cos(\omega t + \varphi^+) \\
    V_{m}^+ \sin(\omega t + \varphi^+) 
    \end{bmatrix}
    \end{equation}
    \end{minipage}
\vspace{+4pt}    
    \begin{minipage}[c]{0.48\textwidth}
    \begin{equation}
    \begin{bmatrix}
        v_{\alpha}^- \\
        v_{\beta}^- 
    \end{bmatrix}
    = T_1
    \begin{bmatrix}
        v_a^- \\
        v_b^- \\
        v_c^- 
    \end{bmatrix}
    = 
    \begin{bmatrix}
    V_{m}^- \cos(\omega t + \varphi^-) \\
    V_{m}^- \sin(\omega t + \varphi^-) 
    \end{bmatrix}
    \end{equation}
    \end{minipage}
\end{subequations}
where
\[
    T_1 =  \dfrac{2}{3}
    \begin{bmatrix}
        1 & -1/2 & -1/2 \\
        0 & \sqrt{3}/2 & -\sqrt{3}/2 \\
        1/2 & 1/2 & 1/2
    \end{bmatrix}
\]
By using the Clarke-Park transformation matrix $T_2$, the voltages in RRF ($dq$-coordinate) are derived:

\begin{subequations}\label{eq:dq}
\centering
\vspace{+3pt}
    \begin{minipage}[c]{0.48\textwidth}
    \begin{equation}
    \begin{bmatrix}
        v_{d}^+ \\
        v_{q}^+ \\
    \end{bmatrix}
    =T_2
    \begin{bmatrix}
        v_{a}^+ \\
        v_{b}^+ \\
        v_{c}^+
    \end{bmatrix}
    =
    \begin{bmatrix}
        V_{m}^- \cos \varphi^+\\
        V_{m}^- \sin \varphi^+
    \end{bmatrix}
    \end{equation}
    \end{minipage}
\vspace{+4pt}    
    \begin{minipage}[c]{0.48\textwidth}
    \begin{equation}
    \begin{bmatrix}
        v_{d}^- \\
        v_{q}^- \\
    \end{bmatrix}
    =T_2
    \begin{bmatrix}
        v_{a}^- \\
        v_{b}^- \\
        v_{c}^-
    \end{bmatrix}
    =
    \begin{bmatrix}
        V_{m}^- \cos(2\omega t + \varphi^-) \\
        -V_{m}^- \sin(2\omega t + \varphi^-)
    \end{bmatrix}
    \end{equation}
    \end{minipage}
\end{subequations}
where
\[
    T_2 = \dfrac{2}{3}
    \begin{bmatrix}
        \cos(\omega t) & \cos(\omega t - 2 \pi / 3) & \cos(\omega t + 2 \pi / 3) \\
        -\sin(\omega t) & -\sin(\omega t - 2 \pi / 3) & -\sin(\omega t + 2 \pi / 3)
    \end{bmatrix}
\]

\subsection{Inner Voltage Controller Design} \label{design}
\begin{figure}[!t]
    \vspace{-8pt}
    \centering
	\includegraphics[width=0.46\textwidth]{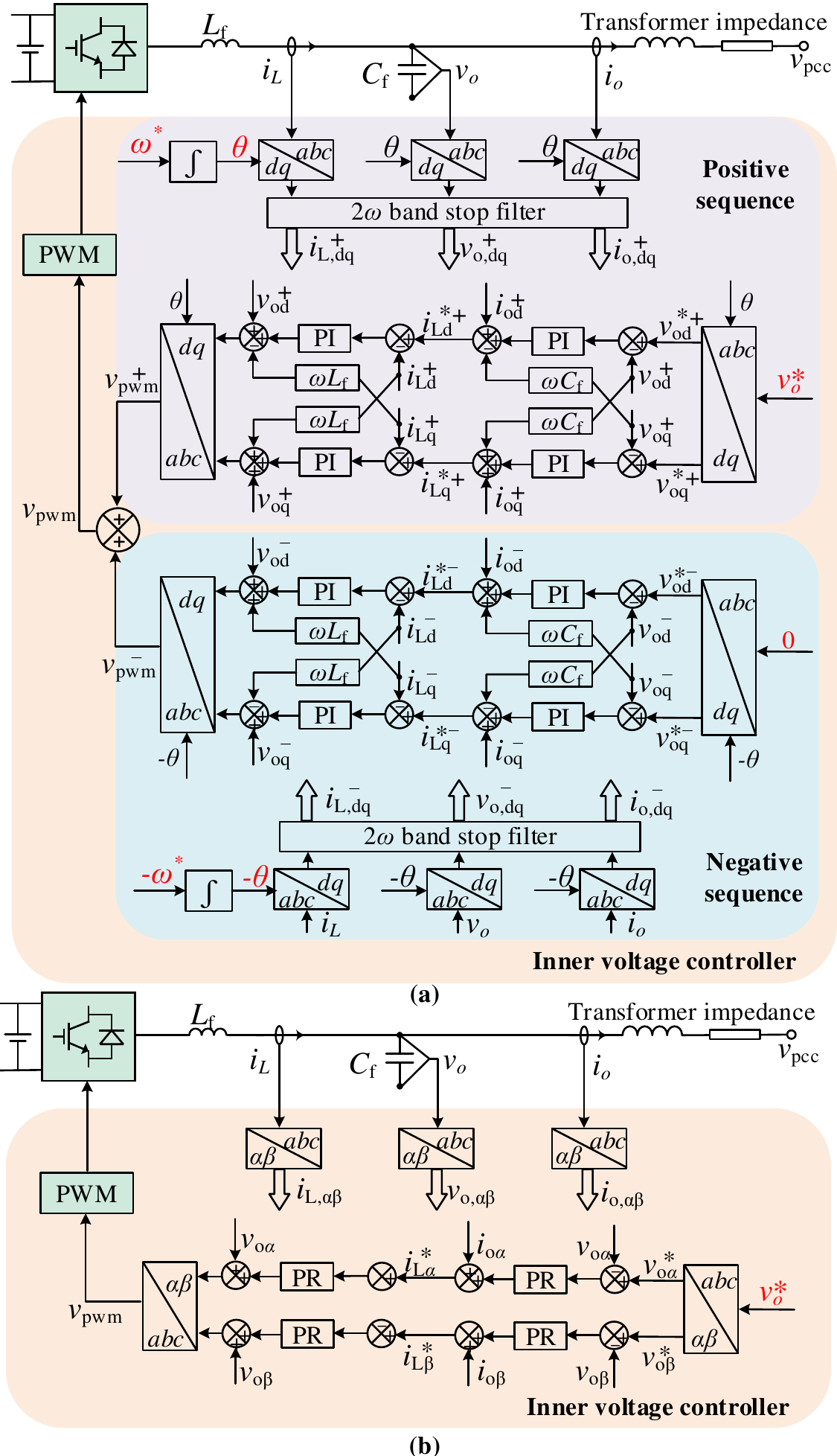}
	\vspace{-7pt}
\caption{Control diagram of the inner controller of a three-phase grid-forming BESS inverter. (a) $dq$ RRF-based control; (b) $\alpha\beta$ SRF-based control.}
\label{fig:control}
% \hfill
\vspace{-7pt}
\end{figure}

Fig. \ref{fig:control}(a) depicts the PS and NS inner voltage controller of the battery inverter in the double synchronously rotating $dq$ reference frames. The three-phase voltages and currents are transformed into PS $dq^+$ axes and NS $dq^-$  axes voltages and currents using a positive phase-locked-loop (PLL) angle $\theta$ and a negative PLL angle $-\theta$, respectively\cite{book_tleis2007power}. 

From \eqref{eq:dq}, it can be seen that although the PS component decomposed to the $dq$ coordinate is DC, the $dq$ projection of the NS component is not DC but a second harmonic ($2\omega$) AC. Thus, monitoring and controlling PS and NS voltages require a dual-reference control scheme, PS $dq^+$ RRF and NS $dq^-$ RRF, rotating in the opposite direction. In addition, the second harmonics should be filtered or removed from the PS and NS $dq$-axes quantities before being sent to the PI controllers in the inner current and voltage control loops. Thus, band-stop filters (notch filters) are used to reject double-frequency components while leaving other frequencies components unaltered. However, adding the filter introduces filter delay and slows down the convergence of the output phase angle.

Fig. \ref{fig:control}(b) illustrates the inner voltage controller in the SRF. From \eqref{eq:albe}, it can be seen that PS and NS components in $\alpha\beta$ coordinate are AC values with the grid's nominal frequency, $\omega=2{\pi}f$. 
Thus, both voltage and current controllers are based on the PR control instead of the conventional PI control. The advantage of the PR control is that it can provide infinite gain at the selected resonant frequency. Therefore, it has excellent tracking capability of AC variables. Because the inner loops are implemented in two-phase stationary frame and all the measured voltage and current are directly transformed from $abc$ to $\alpha\beta$ coordinates, the computational burden is greatly simplified.

%From Section \ref{mechanism}, we can find that the quantities in the SRF are sinusoidal with $\omega$ frequency and  
In Section \ref{simulation_compare}, we will conduct a side-by-side performance comparison to demonstrate the advantage of using the $\alpha\beta$ SRF based control structure.

\subsection{Output Transformer Configuration Design} \label{impact}

\begin{figure}[!t]
 \vspace{-8pt}
  \centering
	\includegraphics[width=0.48\textwidth]{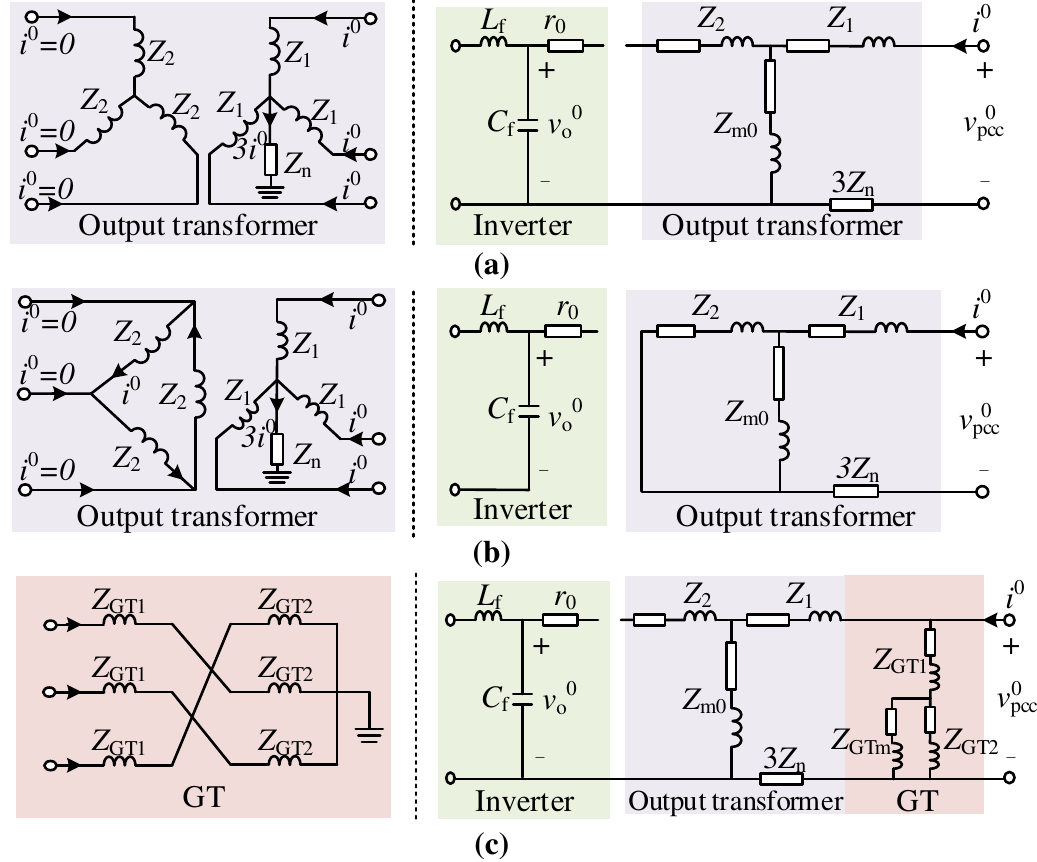}
	\vspace{-7pt}
	\label{fig:transformer_Yg}
    \caption{Zero-sequence equivalent circuit. (a) With Y-Yg output transformer; (b) With $\Delta$-Yg output transformer; (c) With Y-Yg output transformer and GT.}
\label{fig:transformer}
\vspace{-7pt}
\end{figure}

When serving unbalanced loads in power distribution grids, apart from the PS and NS components, ZS components are another contributor to the unbalanced voltage. The ZS circuit of a BESS is determined mainly by the output transformer characteristics, such as the connection method of the primary and secondary windings, the winding grounding arrangements, and the construction type of the magnetic circuit. 

The output transformers for IBRs have two connection types: Y-Yg (see Fig. \ref{fig:transformer}(a)) and $\Delta$-Yg (see Fig. \ref{fig:transformer}(b))\cite{summary_circuit_control}. 
%Thus, in this work, the zero-sequence equivalent circuit of the system with Y-Yg and $\Delta$-Yg transformer are analyzed to find the impact of the output transformer on the voltage unbalance control.
%Fig. \ref{fig:transformer} shows the corresponding zero-sequence equivalent circuits of the Y-Yg or $\Delta$-Yg transformers. 
The ZS impedance of a Y-Yg transformer, $Z_{\mathrm{Y-Y_g}}^0$, and a $\Delta$-Yg transformer, $ Z_{\mathrm{\Delta-Y_g}}^0$, can be expressed as:

\vspace{-7pt}
\begin{equation}\label{eq:Y-Yg}
    Z_{\mathrm{Y-Y_g}}^0 = Z_1 + Z_\mathrm{m0} +3Z_\mathrm{n} \approx Z_\mathrm{m0}
\vspace{-10pt}
\end{equation}

\vspace{-5pt}
\begin{align}\label{eq:delta-Yg}
\begin{split}
    Z_{\mathrm{\Delta-Y_g}}^0  & = Z_1 + (Z_2 // Z_\mathrm{m0}) +3Z_\mathrm{n} \\
    & \approx Z_1 + Z_2 +3Z_\mathrm{n}
\end{split}
\vspace{-7pt}
\end{align}
where $Z_\mathrm{m0} = r_\mathrm{m0}+j\omega L_\mathrm{m0}$ is the transformer magnetizing impedance, $Z_\mathrm{n}$ is the grounding impedance, $Z_1=r_1+j\omega L_1$ and $Z_2=r_2+j\omega L_2$ are the primary and secondary transformer impedance, respectively.

From \eqref{eq:Y-Yg} and \eqref{eq:delta-Yg}, it can be seen that $Z_{\mathrm{Y-Y_g}}^0 \gg Z_{\mathrm{\Delta-Y_g}}^0$ because the magnetizing impedance is thousands of times higher than transformer winding impedance and grounding impedance by design. Thus, the ZS voltage and current in a system with a Y-Yg output transformer would be higher than those with a $\Delta$-Yg output transformer. However, because $\Delta$-Yg transformer configuration provides the additional ZS path, upstream relays might not trip during the ground fault. Thus, $\Delta$-Yg configuration should not be used in most ground-connected distribution networks \cite{ground_vukojevic2020}.

To reduce the ZS voltage at the PCC when using a Y-Yg output transformer, we propose to add a GT ($\Delta$-Yg or Zig-Zag configuration) to the PCC (See Fig. \ref{fig:ciruit}). As shown in Fig. \ref{fig:transformer}(c), by providing the ZS current with a low-impedance path to flow through, the ZS impedance, $Z_{\mathrm{Y-Y_g/GT}}^0$, is:
\begin{align}\label{eq:Y-Yg-GT}
\begin{split}
    Z_{\mathrm{Y-Y_g/GT}}^0  & = (Z_\mathrm{GT1}+Z_\mathrm{GTm}//Z_\mathrm{GT2}) )//(Z_\mathrm{m0}+3Z_\mathrm{n} + Z_1)\\
    & \approx Z_\mathrm{GT1}+Z_\mathrm{GT2}
\vspace{-5pt}
\end{split}
\vspace{-5pt}
\end{align}
where $Z_\mathrm{GT1}$, $Z_\mathrm{GT2}$, $Z_\mathrm{GTm}$ are the primary, secondary, and magnetizing impedance of the GT, respectively. As $Z_{\mathrm{GT}}$ is much smaller than $Z_\mathrm{m0}$, the ZS voltage at the PCC can be significantly reduced. 

From Fig. \ref{fig:transformer}, we can see that ZS current cannot flow in the three-wire ungrounded inverter, so the inverter inner voltage controller cannot sense and regulate the ZS components at the PCC by design. Thus, adding a GT to reduce the ZS voltage at the PCC is the best option for maintaining balanced $v_{\mathrm{pcc}}$.
Note that adding a GT also provides an effective grounding for the BESS during ground faults. 

%\textcolor{blue}{From Dr. Lu: I didn't see how the grounding path can filter the zero sequence current out.  It can only reduce the zero sequence network impedance. The zero-sequence component does not seem to be used by your controller.  This is because in Fig 1, you didn't collect any measurements from PCC (the Yg side of the transformer). So, I think you add the transformer to reduce the impedance in the zero sequence network, which in turn reduce the zero-sequence voltage magnitude measured at the PCC. As Vpcc = V1+V2+V0, this reduce the unbalance of the $V_{pcc}$.}

%\textcolor{blue}{Response: Yes! (1) The GT can only help reduce the zero-sequence voltage while cannot eliminate the zero-sequence voltage. (2) Zero-sequence component cannot be regulated by the inverter controller. Because the inverter can only control its output terminal, $v_o$, so from Fig.3, the zero sequence voltage regulation for $v_o$ cannot be reflected in $v_{pcc}$.}

\section{Simulation Results Analysis} \label{simulation}
To verify the performance of the proposed unbalance voltage control scheme, an electromagnetic transient (EMT) simulation testbed for the IEEE 123-bus distribution feeder system was developed in OPAL-RT/eMEGASIM. A 2-MVA BESS with the topology shown in Fig. \ref{fig:ciruit} is connected at the feeder head of the feeder (i.e., Node 149). The BESS is the only grid-forming resource in the system. To test the BESS operation in two operation conditions: supplying and absorbing unbalanced power, we add a 5-MVA, grid-following PV system (parameters given in \cite{unified2021}) to the feeder at node 149. The electrical and controller parameters of the BESS are listed in Table~\ref{parameter}.
The GT is designed based on IEEE 1547.8 standard\cite{standard1547} (rated at 3.772 MVA with ZS impedance of 0.6185 p.u.).

\begin{table}[!t] 
  \centering
  \vspace{-8pt}
  \caption{Simulation Parameters}\label{parameter}
  \vspace{-3pt}
  \begin{tabularx}{\columnwidth}{p{5.5cm}cc}
    \hline
    \hline
    \multicolumn{2}{c}{\textbf{Inverter Electrical Parameters}}  \\
    \hline
     System Frequency $f$  & 60 Hz  \\
     PCC Nominal Voltage $v_\mathrm{pcc}$ (line-line)   & 4160 $\mathrm{V_{rms}}$ \\
     Inverter Nominal Voltage $v_\mathrm{o}$ (line-line) & 480 $\mathrm{V_{rms}}$\\
     Battery Nominal Voltage $V_\mathrm{b}$ & 1200 V  \\
     LC Filter Inductance $L_\mathrm{f}$ &  350 µH\ \\
     LC Filter Inductance $C_\mathrm{f}$ &  5000 µF \\
    \hline
    \multicolumn{2}{c}{\textbf{Inverter Controller Parameters}}\\
     \hline
     Current PR controller Proportional gain $k_\mathrm{pi}$ & 4\\
     Current PR controller Resonant gain $k_\mathrm{ri}$  & 200\\
     Voltage PR controller Proportional gain $k_\mathrm{pv}$ & 2\\
     Voltage PR controller Resonant gain $k_\mathrm{rv}$  & 1000\\
     Droop controller Coeff. $n_\mathrm{p}$, $n_\mathrm{q}$ & $1\mathrm{e}^{-7}$\\
     Secondary freq. PI controller gain $k_\mathrm{2pf}$, $k_\mathrm{2if}$ & 0.3, 10\\
     Secondary voltage PI controller gain $k_\mathrm{2pv}$, $k_\mathrm{iv}$ & 0.001, 0.5\\
     \hline
     \multicolumn{2}{c}{\textbf{Output Transformer Parameters}}\\
     \hline
     Nominal Power, frequency & 5 MVA, 60Hz\\
     Nominal Voltage (line-line)   & 480 $\mathrm{V_{rms}}$ / 4160 $\mathrm{V_{rms}}$\\
     Resistance $r_\mathrm{1}$ $r_\mathrm{2}$, leakage inductance $L_\mathrm{1}$ $L_\mathrm{2}$ & 0.0012 p.u., 0.03 p.u.\\
     Magnetization impedance $r_\mathrm{m}$, $L_\mathrm{m}$   & 200 p.u., 200 p.u.\\
     Core type  & Three-limb core\\
     \hline
     \hline
   \end{tabularx}
 \vspace{-10pt}
\end{table}

\subsection{Performance Comparison between the SRF and RRF based Controller Design Methods} \label{simulation_compare}
\begin{figure}[!t]
\vspace{-8pt}
\centering
	\includegraphics[width=0.49\textwidth]{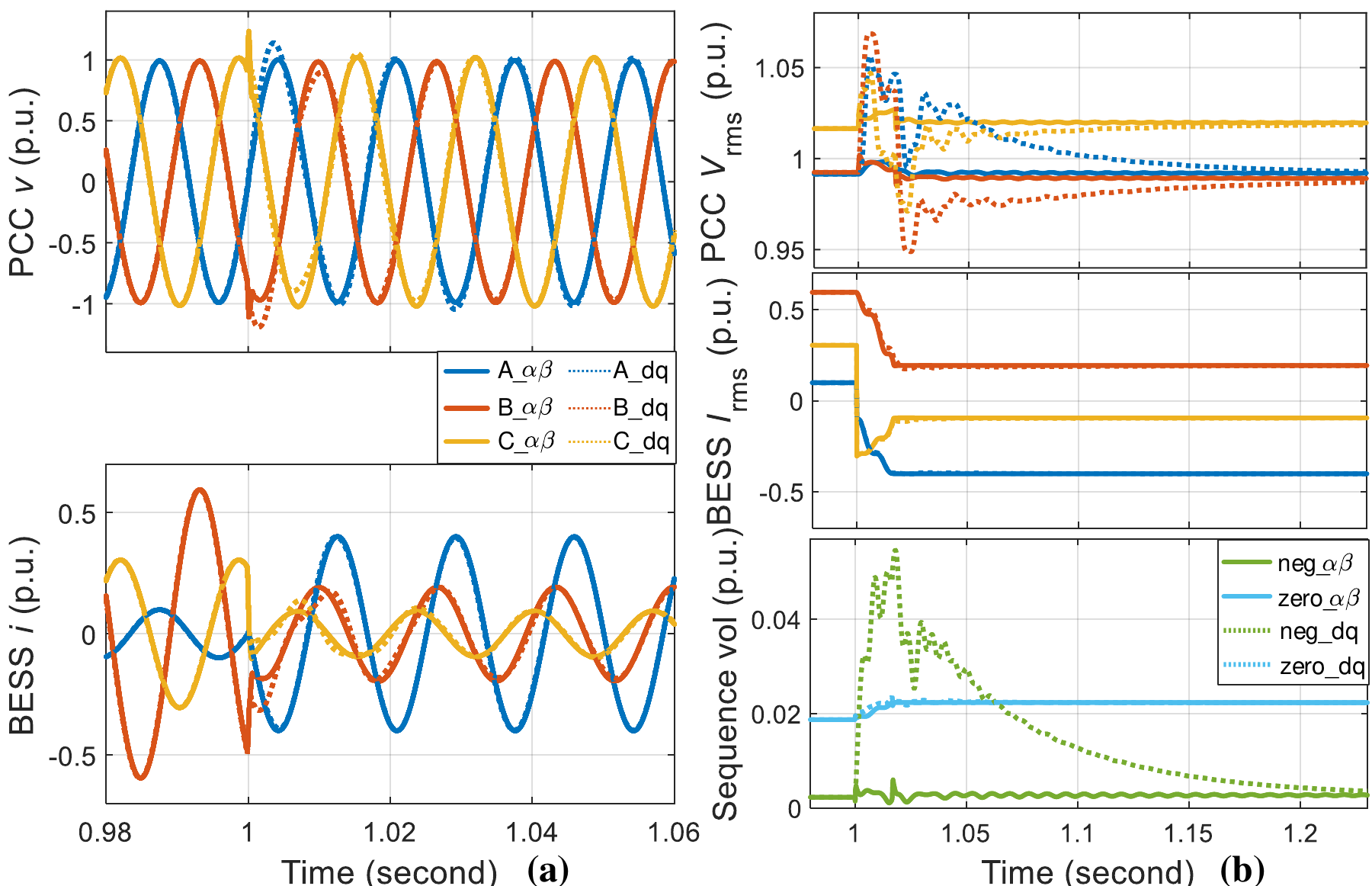}
	\vspace{-15pt}
    \caption{Performance comparison of $\alpha\beta$-based and $dq$-based control methods. (a) Time-series waveforms of PCC voltage and current; (b) RMS profiles of PCC voltage and BESS current, and NS and ZS components of PCC voltage}
\label{fig:resultCompare}	
\vspace{-10pt}
\end{figure}

We compare the controller performance of the proposed $\alpha\beta$ SRF-based method with the conventional $dq$ RRF-based method. %The Y-Yg output transformer as well as the GT are utilized.
As shown in Fig. \ref{fig:resultCompare}, initially, the BESS operates in battery discharging mode with unbalanced output three-phase current ($I_\mathrm{A}=0.1$ p.u., $I_\mathrm{B}=0.6$ p.u., $I_\mathrm{C}=0.3$ p.u.). At $t=1$s, an unbalanced load step change is applied so that the BESS absorbs power from the grid at phase $a$ and $c$ ($I_\mathrm{A}=-0.4$ p.u. and $I_\mathrm{C}=-0.1$ p.u.) and supplies power to the grid at phase $b$ ($I_\mathrm{B}=0.2$ p.u.). 

The steady-state value of PCC voltage and current are the same in both cases, which indicates that both control schemes have a good current tracking performance in steady-state. However, the dynamic performance of the proposed $\alpha\beta$ SRF-based control scheme is significantly better than that of the $dq$ RRF-based method. As shown in Fig. \ref{fig:resultCompare}(c), when taking the RRF-based approach, the NS voltage overshoots and it takes 200 ms to damp. As a result, the PCC voltage can overshoot above 1.05 p.u. and the unbalance of the three-phase voltage is significant in the 200-ms period. The results demonstrate that because no filters are required when using the SRF-based approach, its voltage tracking performance outperforms the RRF-based approach.

\subsection{Power-Voltage Unbalance Curves} \label{simulation_impact}
Voltage unbalance factor (VUF) and power unbalance factor (PUF) are commonly used to quantify voltage and load power unbalance. The definitions of VUF can be found in standards, such as\cite{standard1547,standard1159,standardNEMA}. 
Most standards require electrical supply systems to be designed for limiting the maximum voltage unbalance to 3$\%$.
As introduced in \cite{VUF_definition}, when considering ZS voltage in four-wire system, VUF is calculated as:
\begin{equation}
    VUF = \sqrt{V_-^2+V_0^2}/|V_+|
\end{equation}
where $V_+$, $V_-$, and $V_0$ are the magnitude of PS, NS and ZS voltage.
In this paper, we calculate PUF as:
\begin{equation}
    PUF = \dfrac{Max(|P_\mathrm{A}-P_\mathrm{avg}|,|P_\mathrm{B}-P_\mathrm{avg}|,|P_\mathrm{C}-P_\mathrm{avg}|)}{|P_\mathrm{rated}|}
\end{equation}
where $P_\mathrm{A/B/C}$ represents the power magnitude of phase $a/b/c$,  $P_\mathrm{avg}=(P_\mathrm{A}+P_\mathrm{B}+P_\mathrm{C})/3$ is the three-phase power average, and $P_\mathrm{rated}$ is the rated power per phase.

\begin{figure}[!t]
 \vspace{-10pt}
  \centering
	\includegraphics[width=0.4\textwidth]{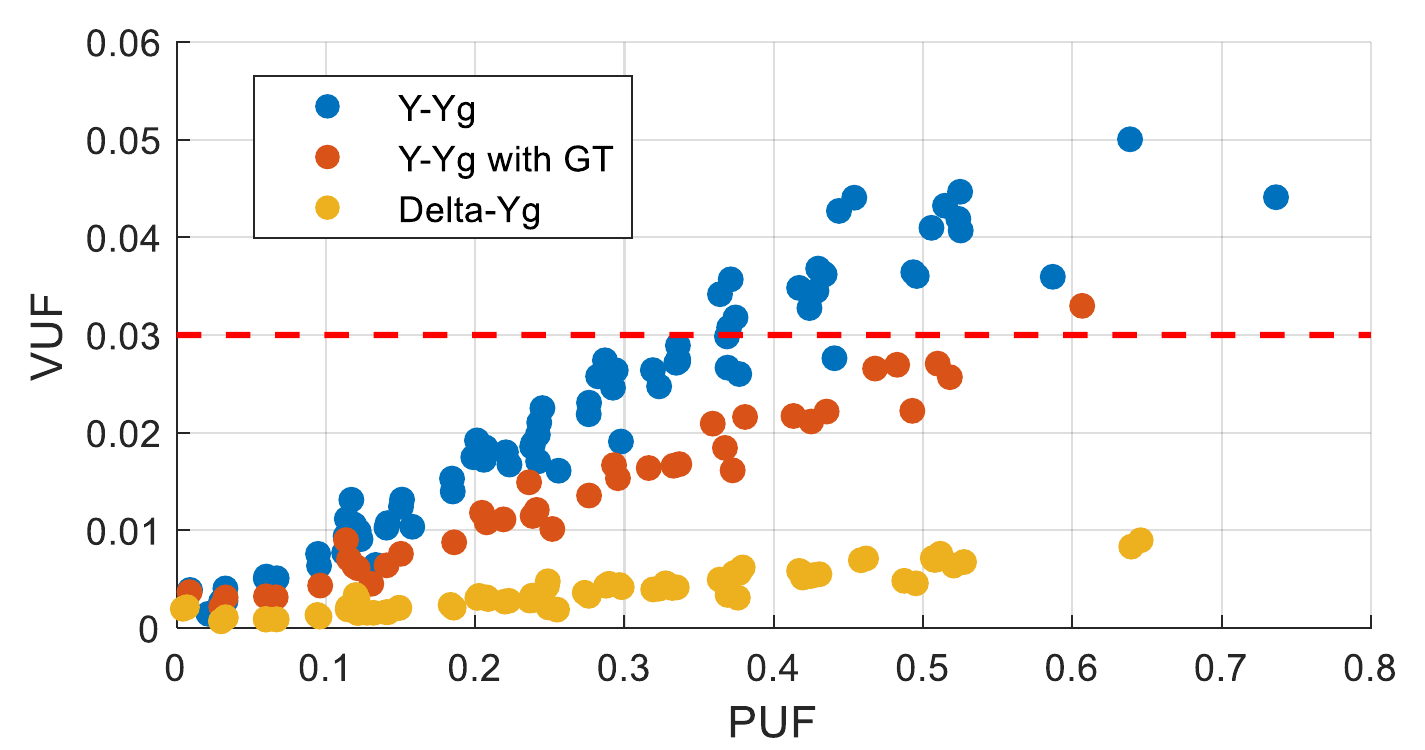}
    \vspace{-10pt}
\caption{PUF versus VUF for different transformer configurations}
\label{fig:PUFVUF}
\vspace{-5pt}
\end{figure}

Experiments can be conducted for different PUF values so that the corresponding VUF points can be computed. Then, a PUF-UVF curve can be obtained, as shown in Fig. \ref{fig:PUFVUF}. 
%1) with Y-Yg output transformer; 2) with Y-Yg output transformer and GT; 3) with $\Delta$-Yg output transformer.
%The results verify the analysis in Section \ref{impact}. 
From the figure, we can see that the VUF of the $\Delta$-Yg output transformer is very linear and when PUF increases from 0 to 0.65, the VUF can be maintained below 1\%. However, if a Y-Yg output transformer is used, VUF can be regulated within 3$\%$ only when PUF is 30$\%$ or less. If a GT is added to the Y-Yg output transformer, VUF can be regulated within 3$\%$ when PUF 55$\%$ or less, a significant improvement to the system capability of supplying unbalanced loads. The importance of the PUF-VUF curve is that it reveals the power unbalance limit, which can be used by an energy management system as an operational constraint for maintaining power balance. 

%this analysis may not be a very fair comparison if the Delta-Yg transformers are not same ratings. Must highlight that the BESS transformer is 5MVA. What if del-yg transformer was also 3.7MVA, would it be an exact match with the GT?
 
\subsection{Real-time Simulation Results} \label{simulation_HIL}

\begin{figure}[!t]
%\vspace{+4pt}
\centering
	\includegraphics[width=0.41\textwidth]{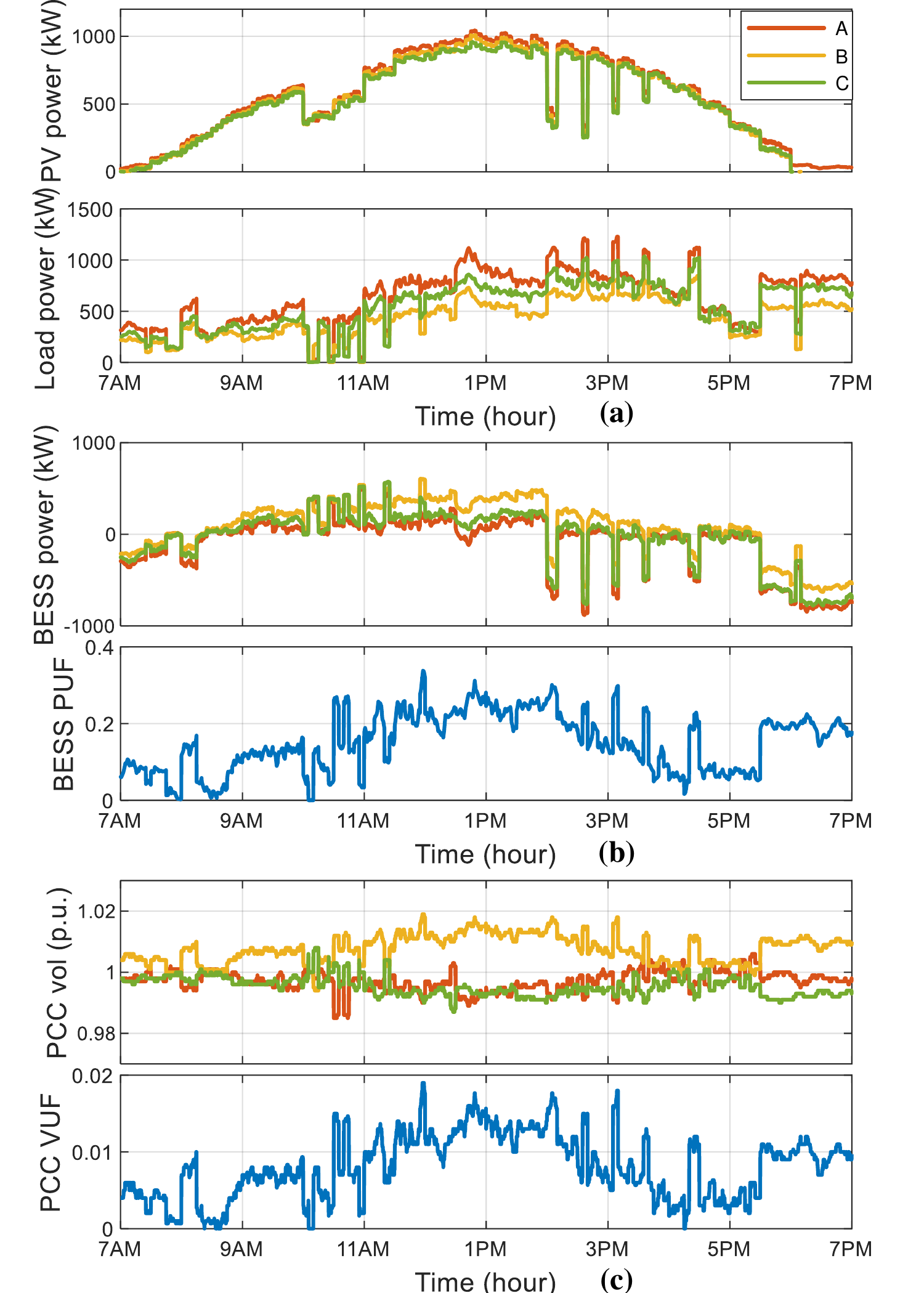}
\vspace{-7pt}
\label{fig:resultHILpower}
\caption{Real-time simulation results for the IEEE 123-bus testbed. (a) PV and load power; (b) BESS output power and PUF; (c) PCC voltage and VUF}
\label{fig:resultHIL}	
\vspace{-12pt}
\end{figure}

The proposed control scheme for mitigating voltage unbalance is also verified on a real-time simulator, OP5700-OPAL-RT, with a 12-hour simulation from 7:00 am to 7:00 pm. The model in OPAL-RT is sampled at every 100us. 
As shown in Fig. \ref{fig:resultHIL}(a), the PV and load power scheduling are based on the energy management algorithm in \cite{hu2020load}, where PV curtailment and load shedding are executed to meet the power supply needs. From Fig. \ref{fig:resultHIL}(b)(c), we can see that VUF is within 3$\%$ if PUF is below  55$\%$.
Note that PUF and VUF have similar shapes. This is because the PUF-VUF curve is close to linear when adding a GT to the Y-Yg output transformer, as shown by the red dots in Fig. \ref{fig:PUFVUF}.

\section{Conclusion} \label{conclusion}
In this paper, we propose an $\alpha\beta$ SRF-based voltage unbalance control architecture for a grid-forming BESS. Compared with the $dq$ RRF-based approach, the proposed method eliminates band-stop filters and achieves superior tracking capabilities. 
%The positive- and negative-sequence voltage can be regulated using its inner voltage controller. 
We propose to include a GT at the PCC to further reduce the system voltage unbalance. We derive the PUF-VUF curves for different transformer configurations. This allows the power unbalanced limits to be calculated from the inverter voltage unbalance limits for microgrid power management purposes. The simulation results demonstrate that when using the proposed SRF-based inner voltage control loop and adding a GT with the Y-Yg connected output transformer, VUF can be regulated within 3$\%$ when the PUF is limited within 55$\%$.
%The proposed control strategy is implemented in an IEEE 123-bus distribution feeder testbed and validated through simulation.

%\section*{Acknowledgment}
\bibliographystyle{IEEEtran}
\bibliography{reference}

\end{document}